\documentclass[12pt]{iopart}
\usepackage{iopams}
\usepackage{graphics}
\begin{document}

\title[Energy Conditions in Palatini Formalism of $f(R)$]{Energy Conditions in Palatini Approach to Modified $f(R)$ Gravity}

\author{H. Saiedi } \ \\
\address{Department of Physics, Florida Atlantic University, FL 33431, USA  }
\ead{\mailto {hsaiedi2014@fau.edu \ ; \  hrssaiedi@gmail.com}}
\begin{abstract}
In this paper, we  review modified $f(R)$ theories of gravity in Palatini formalism. In this framework, , we use the Raychaudhuri's equation along with the requirement that the gravity is attractive, which  holds for any geometrical theory of gravity to discuss the energy conditions. Then, to derive these conditions, we obtain an expression for effective pressure and energy density by considering FLRW metric.  Energy conditions derived in Palatini version of $f(R)$ Gravity differ from those derived in GR. We will see that the WEC (weak energy condition) derived in Palatini formalism has exactly the same expression in its  metric approach. \\  \\

\end{abstract}

\section{Introduction} \
According to astronomical observations, the gravity force at large scales may not behave in standard GR derived from the Hilbert-Einstein action,
$A = \frac{1}{16\pi G}\int {\sqrt{-g} \ R \ d^4x} \ + \int {\sqrt{-g} \ L_m \ d^4x}$, where $R$ is the Ricci scalar, $G$ is Newton's gravitational
constant, and $L_m$ is the matter lagrangian density, respectively [1-4]. So, a generalized Hilbert-Einstein action may be required to fully understand
the gravitational interaction. One of the possible ways to generalize GR is related to the modification of the geometric section of Hilbert-Einstein
action. Examples of such modified gravity models are introduced in [5,6], by assuming that the Ricci scalar $R$ in the lagrangian is replaced by an
arbitrary function $f(R)$ of the Ricci scalar. For discussions of modified $f(R)$ gravity theories see [7-18]. The modified $f(R)$ theories of gravity
can easily explain the recent cosmological observations, and give a solution to the dark matter problem [19]. The metric and Palatini formalisms are two different ways that GR can be derived and lead to the same field equations [20]. However, in modified $f(R)$ gravity, the equations of motion in Palatini approach and metric formalism are generically different [20]. The field equations in metric approach are higher-order while in Palatini approach they are second-order. Both these formalisms in $f(R)$ theories allow the formulation of simple extensions of Einstein's GR. For further discussions of $f(R)$ theories of gravity involving geometry and matter coupling see [21, 22].

In the cosmological context, different $f(R)$ models give rise to the problem of how to constrain from theoretical and observational aspects of these possible $f(R)$ models. Recently, by testing the cosmological viability of some specific cases of $f(R)$ this possibility has been discussed [23-29]. By imposing the energy conditions, we may have further constrains to $f(R)$ theories of gravity [30, 31]. In different contexts, these conditions (so-called energy conditions) have been used to obtain global solutions for a variety of situations. As an example, the weak (WEC) and strong (SEC) were used in the Hawking-Penrose singularity theorems. Also, the null energy condition (NEC) is required to prove of the second law of black hole thermodynamics. However, the energy conditions were basically  formulated in GR [32], one can drive these conditions in $f(R)$ theories of gravity by introducing new effective pressure and energy density defined in Jordan frame. In the present paper, the energy conditions for $f(R)$ theories in Palatini formalism are derived by using the Raychaudhuri's equation (along with the attraction of gravity) which is the ultimate origin of the energy conditions. \\

\section{Palatini Formalism for $f(R)$ Theories of gravity}
To explain the cosmic speed-up, the model $f(R)=R-\mu^4/R$ in metric formalism has some problems. By observation that these problems could be avoided by considering its Palatini formalism, this approach to $f(R)$ theories of gravity has been boosted. Also, the energy conditions in GR and metric formalism of $f(R)$ theories have been discussed in different contexts. So, finding the energy conditions in Palatini version of $f(R)$ would be interesting that we will discuss them in the present paper. Therefore, first we review the field equations in Palatini formalism, and then obtain the energy conditions. The action that defines $f(R)$ theories has the generic form \\
\begin{equation}
A = \frac{1}{2k^2}\int {\sqrt{-g} \ f(R) \ d^4x} \ + A_m[g_{\mu\nu}, \psi_m]
\end{equation} \ \\
where $A_m[g_{\mu\nu}, \psi_m]$ represents the matter action, which depends on the metric $g_{\mu\nu}$ and the matter field  $\psi_m$. In the case of Palatini formalism, the connection $\Gamma^\lambda_{\mu\nu}$ and the metric $g_{\mu\nu}$ are regarded as dynamical variables to be independently varied. Varying the action respect to the metric does yield dynamical equations \\
\begin{equation}
f'(R)R_{\mu\nu}(\Gamma) - \frac{1}{2} f(R) g_{\mu\nu} = k^2 T_{\mu\nu} \ ,
\end{equation} \ \\
where $f'(R)=\frac{df}{dR}$ and $T_{\mu\nu}$ is the usual energy-momentum tensor. $R_{\mu\nu}(\Gamma)$ is the Ricci tensor corresponding to the connection $\Gamma^\lambda_{\mu\nu}$, which is in general different from the Ricci tensor corresponding to the metric connection $R_{\mu\nu}(g)$. Taking the trace of the equation (2), we obtain \\
\begin{equation}
f'(R)R - 2 f(R) = k^2 T \ ,
\end{equation} \ \\
where $R = R(T) = g^{\mu\nu}R_{\mu\nu}(\Gamma)$ is directly related to $T$ and is different from the Ricci scalar $R(g) = g^{\mu\nu}R_{\mu\nu}(g)$ in the metric case. Varying the action (1)  with respect to the connection yields \\
\begin{equation}
\nabla_\alpha(\sqrt{-g} f'(R) g^{\mu\nu}) = 0      \ ,
\end{equation} \ \\
Taking into account that under conformal transformations how the Ricci tensor transforms, it has been shown that [20, 33] \\
\begin{eqnarray}
R_{\mu\nu}(g) - \frac{1}{2} g_{\mu\nu} R(g) &=& \frac{k^2}{f'} T_{\mu\nu} - \frac{R(T)f' - f }{2f'} g_{\mu\nu} + \frac{1}{f'}\left (\nabla_\mu\nabla_\nu f' - g_{\mu\nu}\Box f'\right ) , \ \nonumber \\
&-&  \frac{3}{2(f')^2} \left [ \partial\mu f' \partial\nu f'  -   \frac{1}{2} g_{\mu\nu} (\partial f')^2    \right ]
\end{eqnarray} \ \\
where $R_{\mu\nu}(g)$ and $R(g)$ are computed in terms of the Levi-Civita connection of the metric $g_{\mu\nu}$, i.e., they represent the usual Ricci tensor and scalar curvature. It follows that $R(T) = g^{\mu\nu}R_{\mu\nu}(\Gamma)$ and $R(g) = g^{\mu\nu}R_{\mu\nu}(g)$ are related by \\
\begin{equation}
R = R(T) = R(g) + \frac{3}{2(f')^2} \partial_\lambda f' \partial^\lambda f' - \frac{3}{f'} \Box f' \ \ .
\end{equation} \ \\
For simplicity, we take $k^2 = 8\pi G = 1$. Now, we can realize that the right hand side of equation (5) can be considered as an effective energy-momentum tensor $T^e_{\mu\nu}$. So \\
\begin{eqnarray}
T^e_{\mu\nu} &=& \frac{1}{f'} T_{\mu\nu} - \frac{R(T)f' - f }{2f'} g_{\mu\nu} + \frac{1}{f'}\left (\nabla_\mu\nabla_\nu f' - g_{\mu\nu}\Box f'\right ) , \ \nonumber \\
&-&  \frac{3}{2(f')^2} \left [ \partial\mu f' \partial\nu f'  -   \frac{1}{2} g_{\mu\nu} (\partial f')^2    \right ] \ .
\end{eqnarray} \ \\
Taking the trace of the above equation, one can easily find \\
\begin{equation}
T^e = g^{\mu\nu}T^e_{\mu\nu} =   \frac{T}{f'}  - \frac{2}{f'} \left(R(T)f' - f \right)  -   \frac{3 \Box f'}{f'}   +   \frac{3}{2(f')^2} (\partial f')^2   \ .
\end{equation} \ \\
By substituting $T$ from (3) into equation (8), after simplification,  we reach \\
\begin{equation}
T^e =  \frac{3}{2(f')^2} (\partial f')^2  -  \frac{3 \Box f'}{f'}   -   R(T) \ .
\end{equation} \ \\
Now, by comparing the above relation and equation (6), one can easily realize that \\
\begin{equation}
T^e =  - R(g) \ .
\end{equation} \ \\
So, we can rewrite the equation (5) as \\
\begin{equation}
R_{\mu\nu}(g) = T^e_{\mu\nu} -  \frac{T^e}{2} g_{\mu\nu}   \ ,
\end{equation} \ \\
where $ T^e_{\mu\nu}$ and $T^e$ are  equations (7) and (8), respectively. \\ \\

\section{Energy Conditions in Palatini Version of  $f(R)$ }
To find the energy conditions, we shall use the Raychaudhuri's equation which holds for any geometrical theory of gravity. Therefore, we first briefly review these conditions in GR and then apply them to the $f(R)$ modified gravity in Palatini formalism. The Raychaudhuri's equation implies that for any hypersurface orthogonal congruences, the condition for attractive gravity (convergence of timelike geodesics) reduces to \  $R_{\mu\nu}(g) u^\mu u^\nu \geq 0$, \  where $u^\mu$ is a tangent vector field to a congruence of timelike geodesics. In GR,  using the units such that $k^2 = 8\pi G = c = 1$,  we have  \ $R_{\mu\nu}(g) - \frac{1}{2}g_{\mu\nu}R(g) = T_{\mu\nu}$  \ or \  $R_{\mu\nu}(g) = T_{\mu\nu} - \frac{T}{2}g_{\mu\nu}$. \\
So the condition  $R_{\mu\nu}(g) u^\mu u^\nu \geq 0$ \ implies that \\
\begin{equation}
R_{\mu\nu}(g) u^\mu u^\nu = \left ( T_{\mu\nu} - \frac{T}{2}g_{\mu\nu}   \right ) u^\mu u^\nu \geq 0 \ \ .
\end{equation} \ \\
For a perfect fluid with energy density $\rho$ and pressure $p$
\begin{equation}
T_{\mu\nu} = (\rho + p) u_\mu u_\nu - p g_{\mu\nu} \ .
\end{equation} \ \\
by using the restriction (12), the SEC can be written as \ $(\rho + 3p) \geq 0$. \\
The condition for convergence of null  geodesics along with Einsteins's equations leads to \\
\begin{equation}
R_{\mu\nu}(g) k^\mu k^\nu =  T_{\mu\nu} k^\mu k^\nu \geq 0 \ \ .
\end{equation} \ \\
which is the NEC. Here $k^\mu$ is a tangent vector field to a congruence of null geodesics. Therefore, the NEC for the energy-momentum tensor (13) can be written as \ \ $(\rho + p) \geq 0$. \\

Since, the Raychaudhuri's equation is valid for any geometrical gravity theory, so, the conditions \   $R_{\mu\nu}(g) u^\mu u^\nu \geq 0$ \ and \ $R_{\mu\nu}(g) k^\mu k^\nu  \geq 0$ \ along with  field equations in Palatini version of $f(R)$ gravity implies that \\
\begin{eqnarray}
R_{\mu\nu}(g) u^\mu u^\nu &=& \left ( T^e_{\mu\nu} - \frac{T^e}{2}g_{\mu\nu}   \right ) u^\mu u^\nu \geq 0  , \\  \nonumber \\
R_{\mu\nu}(g) k^\mu k^\nu &=&  T^e_{\mu\nu} k^\mu k^\nu \geq 0 \ \ .
\end{eqnarray} \ \\
where we have used $T^e_{\mu\nu}$ instead of $T_{\mu\nu}$. Now, by comparing the above equations with equations (12) and (14), we can simply figure out that the SEC and NEC can be modified as \ \ $(\rho_e + 3p_e) \geq 0$ \ and \ $(\rho_e + p_e) \geq 0$, respectively. \\
By substituting (6) into (7), for the homogeneous and isotropic Friedmann-Lemaitre-Robertson-Walker (FLRW) metric  with scale factor $a(t)$, and after some simplifications, we reach the following relations for effective density $\rho_e$ and effective pressure $p_e$. \\
\begin{eqnarray}
\rho_e &=& T^{0e}_0 = \frac{1}{f'} \left [  \rho + \frac{1}{2} (f - R(g)f') - \frac{3}{2f'}(\partial_0 f')^2  + \frac{3}{2}\partial_0 \partial_0 f' + \frac{3}{2} H \partial_0 f'   \right ] \ \ ,    \\   \nonumber \\   \nonumber \\
p_e &=& - T^{1e}_1 = \frac{1}{f'} \left [  p - \frac{1}{2} (f - R(g)f')  - \frac{1}{2}\partial_0 \partial_0 f' - \frac{5}{2} H \partial_0 f'   \right ].
\end{eqnarray} \ \\
Here, $H=\dot{a}/a$ is the Hubble parameter. We can easily rewrite the above equations as (by denoting $R_g = R(g)$) \\
\begin{eqnarray}
\rho_e &=& \frac{1}{f'} \left [  \rho + \frac{1}{2} (f - R_gf')    \right ]        \nonumber \\ \nonumber \\
&+& \frac{1}{f'} \left [  \frac{3}{2}\ddot{R_g}f''  - \frac{3}{2f'}\dot{R_g}^2 f''^2   + \frac{3}{2}\dot{R_g}^2 f''' + \frac{3}{2} H \dot{R_g}f''   \right ] \ \ ,  \\ \nonumber \\ \nonumber \\
p_e &=&  \frac{1}{f'} \left [  p - \frac{1}{2} (f - R_gf')  -  \frac{1}{2}\ddot{R_g}f''  -  \frac{1}{2}\dot{R_g}^2 f''' - \frac{5}{2} H \dot{R_g}f''   \right ].
\end{eqnarray} \ \\
To simply express the energy conditions, we can write the Ricci scalar and its derivatives for a spatially flat FLRW metric in terms of the deceleration $(q)$, jerk $(j)$ and snap $(s)$ parameters [34-36] \\
\begin{eqnarray}
R_g &=&   -6H^2(1-q)         \nonumber \\
\dot{R_g} &=&  -6H^3(j-q-2)       \nonumber \\
\ddot{R_g} &=&  -6H^4(s+q^2+8q+6)
\end{eqnarray} \ \\
where \\
\begin{equation}
q=-\frac{1}{aH^2}. \frac{d^2a}{dt^2} \ \ \ , \ \ \ j=\frac{1}{aH^3}. \frac{d^3a}{dt^3} \ \ \ , \ \ \ s=\frac{1}{aH^4}. \frac{d^4a}{dt^4}
\end{equation} \ \\
Now, we can classify the energy conditions as follow \\ \\
\textbf{NEC} \ : \ $(\rho_e + p_e) \geq 0$ \\
\begin{eqnarray}
\rho &+& p   + \ddot{R_g}f''  - \frac{3}{2f'}\dot{R_g}^2 f''^2   + \dot{R_g}^2 f''' +  H \dot{R_g}f'' \ \geq \ 0 \ \ \  \Rightarrow  \nonumber \\ \nonumber \\ \nonumber \\
\rho &+& p   - 6H^4(s+q^2+8q+6)f''  - \frac{54}{f'}H^6(j-q-2)^2 f''^2   \nonumber \\
&+& 36H^6(j-q-2)^2 f''' +6H^4(j-q-2) f'' \ \geq \ 0
\end{eqnarray} \ \\ \\
\textbf{SEC} \ : \ $(\rho_e + 3p_e) \geq 0$ \\
\begin{eqnarray}
\rho &+& 3p -f +R_gf'  - \frac{3}{2f'}\dot{R_g}^2 f''^2   +  6H \dot{R_g}f'' \ \geq \ 0 \ \ \ \Rightarrow \nonumber \\ \nonumber \\ \nonumber \\
\rho &+& 3p - f - 6H^2(1-q)f'     - \frac{54}{f'}H^6(j-q-2)^2 f''^2   \nonumber \\ &+& 36H^4(j-q-2) f'' \ \geq \ 0
\end{eqnarray} \ \\ \\
\textbf{WEC} (week energy condition) \ : \ beside the inequality (23),  \ $\rho_e \geq 0$ \\
\begin{eqnarray}
\rho &+& \frac{1}{2}(f - R_gf')  -  3H \dot{R_g}f'' \ \geq \ 0 \ \ \ \Rightarrow \nonumber \\ \nonumber \\ \nonumber \\
\rho &+& \frac{1}{2}f  + 3H^2(1-q)f'    +  18H^4(j-q-2) f'' \ \geq \ 0
\end{eqnarray} \ \\ \\
\textbf{DEC} (dominant energy condition) \ : \ beside the inequalities (23) and (25), \  $(\rho_e - p_e) \geq 0$ \\
\begin{eqnarray}
\rho &-& p   + f - R_gf'  + 2\dot{R_g}^2 f''' - \frac{3}{2f'}\dot{R_g}^2 f''^2  + 2\ddot{R_g}f''  + 4H \dot{R_g}f'' \ \geq \ 0 \ \ \ \Rightarrow \nonumber \\ \nonumber \\ \nonumber \\
\rho &-& p   + f + 6H^2(1-q) f'  +  72H^6(j-q-2)^2 f''' - \frac{54}{f'}H^6(j-q-2)^2 f''^2 \nonumber \\
&-& 12H^4(s+q^2+8q+6)f''    - 24H^4(j-q-2) f'' \ \geq \ 0
\end{eqnarray} \ \\

It is useful to discuss the energy conditions for some specific $f(R)$ models for the present values of deceleration, jerk, and snap parameters. As we can see from the inequalities (23), (24), (25), and (26), these inequalities depend on the value of the snap parameter except for WEC and SEC. Since a reliable value  of this parameter  has not been reported, therefore, only the WEC and SEC are discussable  with the present observational datas of deceleration and jerk parameters $(q_0=-0.81, \ j_0=2.16)$. We shall note that the inequality (25) for WEC is exactly the same inequality found in [28].  \\

\section{Conclusion} \
In the context of modified $f(R)$ gravity, we have reviewed the field equations in Palatini formalism. It is shown that  the model $f(R)=R-\mu^4/R$ in metric approach  has some problems to explain the cosmic speed-up. By considering the Palatini version of this model, these problems can be avoided.
To discuss the energy conditions, we use the Raychaudhuri's equation along with the requirement that the gravity is attractive, which is the ultimate origin of the energy conditions and holds for any geometrical theory of gravity. We consider FLRW metric to derive the  effective pressure and energy density, which are needed to find the energy conditions.  It is worth to mention here that the WEC derived in Palatini formalism of $f(R)$ gravity is  exactly the same WEC found in its metric approach.

\end{document}